\documentclass[aps,prl,groupedaddress,superscriptaddress,twocolumn]{revtex4-1}
\usepackage{graphicx}
\usepackage{amsmath}
\usepackage{amssymb}
\begin{document}
\title{Superconducting phase transitions in ultrathin TiN films}

\author{T.\,I. Baturina}
\affiliation{ A. V. Rzhanov Institute of Semiconductor Physics SB RAS, 13 Lavrentjev Avenue, 
Novosibirsk, 630090 Russia}
\affiliation{Materials Science Division, Argonne National Laboratory, Argonne, Illinois 60439, USA}
\author{S.\,V. Postolova}
\affiliation{ A. V. Rzhanov Institute of Semiconductor Physics SB RAS, 13 Lavrentjev Avenue, 
Novosibirsk, 630090 Russia}
\author{A.\,Yu. Mironov}
\affiliation{ A. V. Rzhanov Institute of Semiconductor Physics SB RAS, 13 Lavrentjev Avenue, 
Novosibirsk, 630090 Russia}
\author{A.~Glatz}
\affiliation{Materials Science Division, Argonne National Laboratory, Argonne, Illinois 60439, USA}
\author{M.\,R.~Baklanov}
\affiliation{IMEC Kapeldreef 75, B-3001 Leuven, Belgium}
\author{V.\,M. Vinokur}
\affiliation{Materials Science Division, Argonne National Laboratory, Argonne, Illinois 60439, USA}

\date{\today}

\begin{abstract}
We investigate transition to the superconducting state in the ultrathin ($\leq 5$\,nm thick) 
titanium nitride films on approach to superconductor-insulator transition.  
Building on the complete account of quantum contributions to conductivity, we demonstrate
that the resistance of thin superconducting films exhibits a non-monotonic temperature behaviour due to 
the competition between weak localization, electron-electron interaction, 
and superconducting fluctuations. 
We show that superconducting fluctuations give rise to an appreciable decrease in the resistance
even at temperatures well exceeding the superconducting transition temperature, $T_c$, with
this decrease being dominated by the Maki-Thompson process.
The transition to a global phase-coherent superconducting state occurs via the 
Berezinskii-Kosterlitz-Thouless (BKT) transition, which we observe both 
by power-law behaviour in current-voltage characteristics and by 
flux flow transport in the magnetic field. 
The ratio $T_{BKT}/T_c$ follows the universal Beasley-Mooij-Orlando relation.
Our results call for revisiting the past data on superconducting transition in thin 
disordered films.
\end{abstract}


\maketitle
It has long been known that in thin films the
transition into a superconducting state occurs in two stages:
first, with the decreasing temperature, the finite amplitude of the order parameter 
forms at the superconducting critical temperature $T_c$,
second, a global phase coherent state establishes at lower temperature $T_{BKT}$ 
[the temperature of the Berezinskii-Kosterlitz-Thouless (BKT) transition]
~\cite{Beasley1979,HalperinNelson1979}.  
Neither of these two temperatures results in singularities in the temperature 
dependences of the linear resistivity $R(T)$ and determination of $T_c$
and therefore $T_{BKT}$ from the experimental data is a non-trivial task.  
Fortunately, $T_c$ can be determined by juxtaposing measured $R(T)$ with 
the results of the theory of superconducting fluctuations (SF) in the  region $T>T_c$, as
it enters the theory as a parameter (see Ref.~\cite{LarVarBook} for a review). 
$T_{BKT}$ can be determined either by the analysis of the power-law behaviour of
current-voltage ($I$-$V$)  characteristics at zero magnetic field 
or by the change of the curvature of the magnetoresistance 
from the convex to the concave in the magnetic field perpendicular to the film 
 plane~\cite{Minnhagen1981,MinnhagenRevModPhys,Epstein1981,Wolf1981,Kadin1983,
HebardFiory1983,Fiory1983,HebardPaalanen1985,Simon1987}.

Although superconductivity in ultrathin films 
has long been a subject of active investigations, only in the present work became it possible 
to comprehensively take into account 
all quantum contributions to conductivity (QCC).
Our analysis is based on the recent theoretical advance~\cite{GlatzVarVin} 
offering a complete description of fluctuation superconductivity, which is valid 
at all temperatures above the superconducting transition.
We demonstrate, in particular, that the omission of the Maki-Thompson contribution 
leads to the incorrect values of $T_c$.
Our results thus call for revisiting the previously published
data on the superconducting transition in thin disordered films.

The data presented below are taken for thin ($\leq 5$\,nm) TiN films formed 
on a Si/SiO$_2$ substrate by atomic layer deposition and fully
characterized by high-resolution electron beam, infra-red~\cite{OpticsTiN}, 
and low-temperature scanning tunnelling spectroscopy~\cite{STM_TiN,PG_TiN}. 
The films are identical to those that experienced the superconductor-insulator transition 
 (SIT)~\cite{SITTiN,TiN_HA,VinNature} and have
the diffusion constant $D\gtrsim 0.3$\,cm$^2$/sec and the superconducting coherence 
length $\xi_d(0)\gtrsim 8$\,nm.
The samples were patterned into bridges 50\,{$\mu$m} wide and 250\,{$\mu$m} long.  
Transport measurements are carried out 
using low-frequency ac and dc techniques in a four probe configuration.  

We start with the fundamentals of the theory of quantum contributions to conductivity.
A notion of quantum corrections to conductivity can be traced back to the 
work~\cite{LangerNeal1966}, where small corrections to the diffusion coefficient were discussed.  
In the years that followed a breakthrough in understanding the nature and role 
of quantum effects 
was achieved~\cite{Strongin1968,AslLar1968,Maki1968,Thompson1970}, 
crowned by formulation of the concept
of quantum contributions to conductivity in disordered 
metals~\cite{GLKh1979,AslVar1980,LarkinMT1980,AAL1980}.
A theory of QCC (see for review~\cite{LarVarBook,AAreview}) rests on two major phenomena:
First, the diffusive motion of a single electron is accompanied 
by quantum interference of the electron waves [weak localization (WL)] 
impeding electron propagation. 
Second, disorder enhances the Coulomb interaction.
The corresponding 
QCC comprise, in their turn, two components: 
the interaction between particles with close momenta [diffusion channel (ID)] 
and the interaction of the electrons with nearly opposite momenta (Cooper channel).
The latter contributions are referred to as superconducting fluctuations (SF) 
and are commonly divided into four distinct types: 
(i) Aslamazov-Larkin term (AL) 
describing the flickering short circuiting of conductivity 
by the fluctuating Cooper pairs; 
(ii) Depression in the electronic density of states (DOS) 
due to drafting the part of the normal electrons to form Cooper pairs above $T_c$;
(iii) The renormalization 
of the single-particle diffusion coefficient (DCR)~\cite{GlatzVarVin}; 
and last but not least (iv) the
 Maki-Thompson contribution (MT)~\cite{Maki1968,Thompson1970} 
coming from coherent scattering of electrons forming Cooper pairs on impurities. 

The total conductivity of the disordered system is thus the sum of all 
the above contributions  added to the bare Drude conductivity $G_0$:
\begin{equation}
G = G_0+\Delta G^{WL}+\Delta G^{ID}+\Delta G^{SF}\,,
\label{totalG2}
\end{equation}
\begin{equation}
\Delta G^{SF} = \Delta G^{AL}+\Delta G^{DOS}+\Delta G^{DCR}+\Delta G^{MT}\,.
\label{SFtotal}
\end{equation}
These contributions have their inherent temperature and magnetic field behaviours that 
govern the transport properties of disordered systems 
and strongly depend on dimensionality. 
We will focus on quasi-two-dimensional (quasi-2D) systems since our films fall into this category.  
In films the notation $G$ in the Eq.\,\eqref{totalG2} refers to the conductance rather than to conductivity.
The system is quasi-2D if its thickness, $d$, is larger than both, 
the Fermi wavelength, $\lambda _F$, and the mean free path, $l$, 
but is less than the phase coherence length, $l_\varphi$, (responsible for WL effect) and 
the thermal coherence length responsible 
for the electron-electron interaction in both the Cooper channel and the diffusion channel 
($l_T=\sqrt{2\pi \hbar D/(k_BT)}$), 
that is, $\lambda _F,\,l<d< l_\varphi, \, l_T$.
In this case the WL and ID corrections can be written as:
\begin{equation}
\Delta G^{WL}+\Delta G^{ID}=G_{00}A \ln\left[k_BT \tau/\hbar \right]\,,
\label{WLIDG}
\end{equation}
\begin{equation}
A=a p+A_{ID}\,, 
\label{AWLID}
\end{equation}
where $G_{00}=e^2/(2\pi ^2 \hbar)$,
$a=1$ provided the spin-orbit scattering is neglected ($\tau_\varphi \ll \tau_{so}$),
$a=-1/2$ when scattering is relatively strong ($\tau_\varphi \gg \tau_{so}$),
$p$ is the exponent in the temperature dependence of the phase coherence time
$\tau_\varphi \propto T^{-p}$, and $A_{ID}$ is a constant depending on the Coulomb screening
and which in all cases remains of the order of unity~\cite{Finkelstein1983}. 
At low temperatures where electron-electron scattering dominates,
\begin{equation}
\tau_\varphi^{-1}=\frac{\pi kT}{\hbar}\frac{e^2R}{2\pi ^2 \hbar}
\ln\frac{\pi \hbar}{e^2R}\,,
\label{tNy}
\end{equation} 
yielding $p=1$, $R$ is the sheet resistance. 
At higher temperatures where the electron-phonon interaction
becomes relevant, $p=2$.  
This concurs with experimental observations~\cite{Haesendonck1983,Gershenzon1983,
Raffy1983,Santhanam1984,Bergmann1984,Gordon1984,Gordon1986,Brenig1986,WuLin1994}
where $1\leq p\leq 2$, with $p=1$ at $T<10$\,K.

\begin{figure}[t]
\includegraphics[width=1.0\columnwidth]{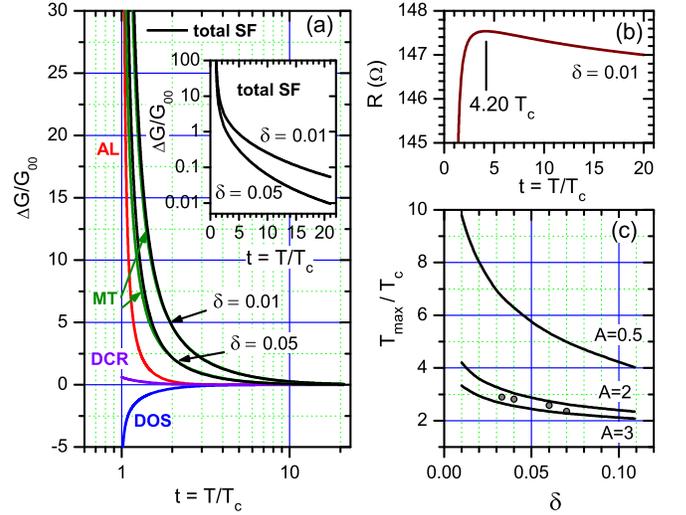}
\caption{\label{fig:TheorySF}
 (Color online)
(a) Temperature dependences of superconducting fluctuation contributions to conductivity 
[in the units of $G_{00}=e^2/(2\pi ^2 \hbar)$]~\cite{GlatzVarVin}.
The curves for AL, DOS, and DCR processes are universal functions of 
reduced temperature $t=T/T_c$, 
the MT contribution is presented for $\delta=0.01$ and $0.05$.
The black solid lines are the sum of all SF contributions~Eq.\,\eqref{SFtotal}.  
The inset shows the same total SF contribution on logarithmic scale. 
(b) Resistance vs. reduced temperature (see details in the text).
(c)  The set of the curves $T_{max}/T_c$ vs. $\delta$ 
for different coefficients $A$ from Eq.\,\eqref{WLIDG}.  
The circles represent the measured $T_{max}$ and $T_c$ and $\delta$ 
obtained by fitting the experimental data (see the Table I and discussion in the text).
}
\end{figure}

As far as superconducting fluctuations are concerned, the complete comprehensive formula, which
includes all the SF contributions and is valid at all temperatures and magnetic fields 
above the superconducting transition line $B_{c2}(T)$, 
was recently derived in Ref.~\cite{GlatzVarVin}.  
We do not reproduce here this somewhat cumbersome expression, but present the results 
of the calculations in the zero magnetic field in Fig.\,\ref{fig:TheorySF}.  
The AL, DOS, and DCR contributions
are universal functions of the reduced temperature $t=T/T_c$.   
The corresponding tabulated expressions are plotted in Fig.\,\ref{fig:TheorySF}a.  
The peculiarity of the MT contribution is that it depends on
the phase coherence time $\tau_{\varphi}$, 
which enters through the parameter 
$\gamma_{\varphi}=\pi\hbar/(8k_{\scriptscriptstyle B}T_c\tau_{\varphi})$.  
The latter can be expressed through the conventional pair-breaking parameter 
\begin{equation}
\delta=\pi\hbar/(8k_BT\tau_\varphi)
\label{delta1}
\end{equation} 
as $\gamma_{\varphi}=t\delta $.  
Note, that if $\tau_{\varphi}\propto T^{-1}$, 
then $\delta$ becomes temperature independent and by using the expression for 
$\tau_\varphi$ from Eq.\,\eqref{tNy}, Eq.\,\eqref{delta1} can be rewritten as
	\begin{equation}\label{deltaR}
		\delta=e^2R/(16 \hbar)\ln[\pi \hbar/(e^2R)]\,.
	\end{equation}
The MT contributions for the two values of the parameter $\delta=0.01$ and $\delta=0.05$ 
are presented in Fig.\,\ref{fig:TheorySF}a.

The following comments are in order.  
The calculations clearly demonstrate that at temperatures $T> 2T_c$ 
the \textit{total} fluctuation-induced conductivity with the great precision 
merely coincides with that given by the MT term.  
In other words, in this temperature range the sum of contributions 
$\Delta G^{AL}+\Delta G^{DOS}+\Delta G^{DCR}$=0.  
Moreover, the MT process dominates the fluctuation superconductivity down to temperatures 
$t-1\lesssim\delta$.  
At lower temperatures the AL contribution starts to become larger.  
The fact that MT is a leading process in a wide range of temperatures 
has already been emphasized in very early works by Maki~\cite{Maki1968} 
and Thompson~\cite{Thompson1970} and subsequent 
theoretical works~\cite{AslVar1980,LarkinMT1980}.  
On the experimental side, numerous experimental studies 
that demonstrate that the magnetoresistance at $T>T_c$ is caused mainly by 
the suppression of the MT 
process~\cite{Haesendonck1983,Gershenzon1983,Raffy1983,Santhanam1984,
Bergmann1984,Gordon1984,Gordon1986,Brenig1986,WuLin1994,MTPtSi}
support this conclusion.

Next, it is the common view that the temperature range $(T-T_c)/T_c\equiv t-1$ where 
superconducting fluctuations are relevant is defined by the 
so-called Ginzburg-Levanyuk parameter, $t-1\lesssim$Gi.
Note that in the two-dimensional case Gi$=e^2R/(23\hbar)$ 
and as follows from \eqref{deltaR}  Gi$\approx\delta$. 
Therefore Gi defines only a narrow vicinity of $T_c$ where the AL term becomes dominant.
What concerns the total SF contribution, it remains noticeable and positive even 
at temperatures well above $T_c$, as it is clearly demonstrated 
by the inset in Fig.\,\ref{fig:TheorySF}a presenting the total SF contribution on logarithmic scale.

As the measurable quantity is the resistance, rather than the conductance, 
it is instructive to recast the calculated QCC  
into the temperature dependence of the resistance, $R(T)$.
To this end, we use Eq.\,\eqref{deltaR} which offers a reasonable estimate 
for $G_0=R^{-1}$ in Eq.\,\eqref{totalG2}.
In Fig.\,\ref{fig:TheorySF}b we show the temperature dependence of the resistance calculated as 
$R(t)=[\Sigma\Delta G^{(i)}(t)+G_0]^{-1}$ for  
$\delta=0.01$ yielding $G_0^{-1}=147\,\Omega$. 
We choose the temperature reference point at $T=20T_c$ assuming 
that all the contributions are practically zero at this temperature.
Summing up all the QCC we took the coefficient $A=2$ in Eq.\,\eqref{WLIDG}.
One sees that although the SF contributions alone would have resulted in the monotonic 
behaviour of the resistance (with $dR/dT>0$), 
the contributions from WL and ID processes 
make $R(T)$ become non-monotonic and exhibit a maximum at the some temperature, $T_{max}$, 
of about of few $T_c$.

In Fig.\,\ref{fig:TheorySF}c we plot the ratio $T_{max}/T_c$ as function of $\delta$ 
for three most common experimental
situations where $A=3,\,2$ and $0.5$, corresponding to three sets of parameters 
$(a,p,A_{ID})$ in Eqs.\,\eqref{WLIDG} and \eqref{AWLID}: (1,2,1), (1,1,1), and (-1/2,1,1).
The immediate important conclusion to be drawn from these plots 
is that the maximum in the $R(T)$ dependences for thin superconducting films is always present, 
even for the smallest  values of the coefficient $A$.
However,  in relatively thick films this maximum can be disguised by the classical linear 
drop of the resistance with decreasing temperature.
The next observation is that for all the realistic parameters of the film, 
the larger the resistance $R$ [i.e. the larger $\delta$ Eq.\,\eqref{deltaR}], 
the closer $T_{max}$ moves to $T_c$, maintaining, nevertheless, that  
the ratio $T_{max}/T_c$ remains always larger than 2.
It is noteworthy that the maximum lies in the domain where the SF are dominated 
by the Maki-Thompson contribution and that the maximum itself arises from 
the competition between the WL+ID and MT processes.
In general, $T_{max}/T_c$ vs. $\delta$ curves relate the quantity 
$T_{max}$ which is the only characteristic point in the $R(T)$ dependence with
the transition temperature $T_c$ and as such can serve as a set of calibrating curves 
for the express-determination of $T_c$, since $A$ can be 
estimated from the analysis of the resistance behaviour at high temperatures.

\begin{figure}[tbh]
\includegraphics[width=1.0\columnwidth]{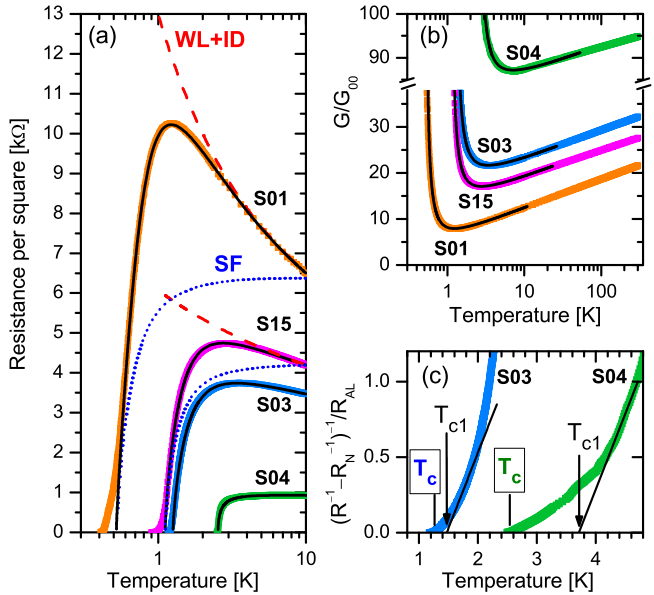}
\caption{\label{fig:RTexp} 
(Color online) 
(a) Resistance per square vs. temperature for four 
TiN film samples listed in Table I.
Solid lines: fits accounting for all the corrections.
Dashed lines (marked as WL+ID): separate contribution of the sum 
of weak localization and interaction in the diffusion channel 
to the resistance of the samples S01 and S15.  
Dotted lines (SF): contribution of superconducting fluctuations.  
(b) The same data as in (a), but extended to room temperatures 
and re-plotted as the dimensionless conductance $G/G_{00}$.
The semi-logarithmic scale representation reveals 
logarithmic decrease of the conductance with 
temperature owing to WL and ID effects.
(c) Reduced conductance  
$(R^{-1}-R_{N}^{-1})^{-1}/R_{AL}$ ($R_{AL}^{-1}=e^2/(16 \hbar)$ and $R_N=R_{max}$)
vs. $T$.  
The linear fit to the AL expression (solid lines) are often used 
for the determination of $T_c$ and generally gives incorrect (much too high) 
values of the critical temperature: $T_{c1}=1.470$\,K, and $3.714$\,K for samples S03 and S04, 
respectively (marked by arrows).
The correct values of $T_c$ (presented in Table I) are in the rectangles 
and marked by the vertical bars.
}
\end{figure}

Now we turn to discussion of our experimental results.  
Figure\,\ref{fig:RTexp}a presents the temperature dependences of the resistance per square 
for four TiN films with different room temperature resistances.  
In all samples the resistances first grow upon decreasing the temperature from 
 room temperature down, then reach the maximum value, $R_{max}$, 
at some temperature $T_{max}$ (see Table I), and, finally, decreases, 
with $T_{max}$ being approximately three times larger
than the temperature where $R$ becomes immeasurably small.
Before fitting the data with the theory of QCC, 
let us verify that the films in question are
indeed quasi-two-dimensional with respect to the effects of the electron-electron interaction. 
We find $l_T\approx 2.5$\,nm at $T=300$\,K
($l_T>12$\,nm at $T=10$\,K).  
Therefore the condition of quasi-two-dimensionality, $d\lesssim l_T$, 
is satisfied at all temperatures down from 300\,K.
That is why the temperature behaviour of the conductance follows the logarithmic temperature 
dependence in accord with Eq.\,\eqref{WLIDG}, see Fig.\,\ref{fig:RTexp}b.
Solid lines in Fig.\,\ref{fig:RTexp}a,b account for all quantum contributions. 

\begin{table}[t]
\caption{Sample characteristics:
$R_{300}$ is the resistance per square at $T=300$\,K;
$R_{max}$ is the resistance at the maximum point achieved at temperature $T=T_{max}$; 
$T_c$ is the critical temperature determined from the quantum contribution fits 
together with the pair-breaking parameter $\delta$ and the coefficient $A$ in Eq.\eqref{WLIDG};
$T_{BKT}^{(I)}$ is determined from the power-law behavior of $I$-$V$ characteristics and
$T_{BKT}^{(B)}$ is found from the flux flow resistance data.
Note that the sample S01 is the same sample \#1 in ref.~\cite{SITTiNJETPL}, where 
the higher $T_c=0.6$\,K was found since only the AL contribution was taken into account. 
}
\label{tab}
\begin{center}
\begin{tabular}{lccccccccccccccccc}
\hline \hline
Film & $R_{300}$&  $R_{max}$& $T_{max}$ & $T_c$ & $\delta$ & A    & $T_{BKT}^{(I)}$ &$T_{BKT}^{(B)}$ \\
$\#$ & k$\Omega$& k$\Omega$ &   K       &  K   &          &      &  K              &K               \\
\hline
S04  &0.855    & 0.932     & 7.34      & 2.538 & 0.033    & 2.63 & 2.497           &   2.475      \\
S03  &2.52      & 3.74      & 3.55      & 1.260 & 0.040    & 2.63 & 1.147           &    1.115      \\
S15  &2.94      & 4.74      & 2.88      & 1.115 & 0.060    & 2.59 & 0.910           &    0.895      \\
S01  &3.75      & 10.25     & 1.23      & 0.521 & 0.070    & 2.71 & ---             &    0.380      \\
\hline \hline
\end{tabular}
\end{center}
\end{table}

The fitting remarkably captures all major features of the observed dependences:
their non-monotonic behaviour, the position and the height of $R_{max}$, 
and the graduate decrease in the resistance 
matching perfectly the experimental points down to values $R\ll R_{max}$
(without any additional assumptions about mesoscopic 
inhomogeneities~\cite{IoffeLarkin1981,Castellani2011}).
We were using three fitting parameters, $\delta$, $A$, and, $T_c$ 
(the values providing the best fit are given in the Table 1).  
It is noteworthy that while varying $\delta$ and $A$ significantly shifts 
the temperature position and the very value of $R_{max}$, 
it does not change noticeably the position of $T_c$.
It reflects the fact that $\Delta G^{SF}$ does not depend on 
the pair-breaking parameter $\delta$ in the close vicinity of $T_c$ 
(see inset to Fig.\,\ref{fig:TheorySF}a where the curves for different $\delta$ merge).
A cross-check of the validity of the extracted values of $\delta$ and $A$ is achieved by
taking the values of $T_{max}/T_c$ and $\delta$ found from fitting and superposing 
them on Fig.\,\ref{fig:TheorySF}c.
We see that the points fall between the lines corresponding to $A=2$ and $A=3$ 
in a nice accord with $A\simeq 2.6$ obtained from the full description of $R(T)$.

Having completed the full description of the experimental data within the framework 
of a general theory of superconducting fluctuations~\cite{GlatzVarVin},
it is instructive to review the approaches for inferring $T_c$ 
from the experimental data that were frequently used in the past.
First, we find that $T_c$ lies at the foot of the $R(T)$ curve where 
$R(T)\simeq(0.08\div 0.13)R_{max}$.  
Therefore, the determination of $T_c$ as the temperature where $R(T)$ drops 
to 0.5$R_N$ (let alone to 0.9$R_N$) significantly overestimates $T_c$.  
Another frequently used procedure~\cite{Fiory1983} is based on the assumption 
that the effect of quantum corrections can be reduced to the AL term only,
i.e. that the resistance obeys the relation $R^{-1}=R_{N}^{-1}+R_{AL}^{-1}/(T/T_c -1)$, 
where $R_{AL}^{-1}=e^2/(16 \hbar)=1.52 \cdot 10^{-5}\,\Omega ^{-1}$.  
This implies that there would have existed the range of temperatures near $T_c$
where the plot $[(R^{-1}-R_{N}^{-1})^{-1}/R_{AL}]$ vs. $T$ could have been approximated 
with a straight line with the slope $=1$.
The intersection of this line with the $T$-axis would have defined $T_c$.
Utilizing this approach we plotted in Fig.\,\ref{fig:RTexp}c the data for two of our samples,
as an example, and indeed find such a linear dependence for each sample
(shown by the solid lines), yielding temperatures of the intersections 
marked as $T_{c1}$.
One sees, however, that this procedure gives much too high values 
for the superconducting critical temperatures. 

\begin{figure}[tbh]
\includegraphics[width=1.0\columnwidth]{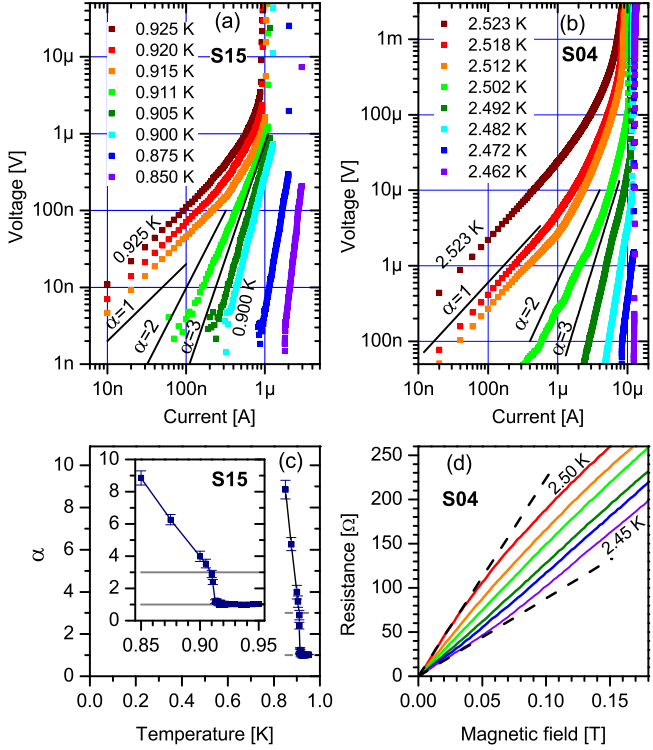}
\caption{\label{fig:VBKTexp} 
(Color online)
Current-voltage characteristics on a double-logarithmic scale 
for samples S15 (a) and S04 (b) at different respective
temperatures listed in the panels.  
Solid lines indicate the slopes corresponding to different values of power $\alpha$ on the 
$V \propto I^{\alpha}$.
(c) $\alpha(T)$ for sample S15 in a full-temperature scale showing that 
the development of the power-law behaviour occurs over a very narrow temperature interval.
The inset presents the same data on a magnified temperature scale, emphasizing the jump 
from $\alpha=1$ to $\alpha=3$ at the transition.
(d) Resistance isotherms vs. magnetic field for sample S04 
measured at temperatures 10\,mK apart. The dashed lines are the tangents to 
lowest and highest isotherms stressing the evolution from 
superlinear to sublinear behaviour. }
\end{figure}

\begin{figure}[tbh]
\includegraphics[width=1.0\columnwidth]{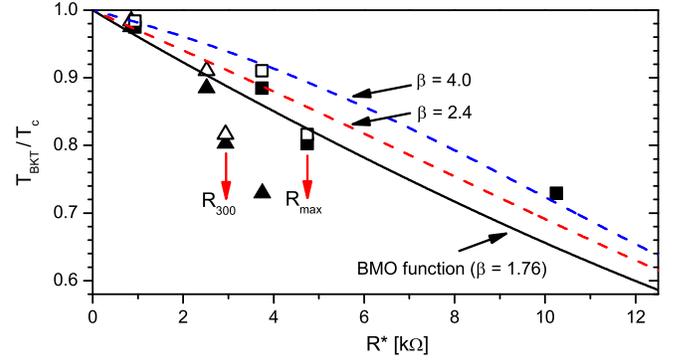}
\caption{\label{fig:TBKTtoTcexp} 
(Color online) Plot of $T_{BKT}/T_c$ as a function of $R^*$ for TiN films.  
The triangles correspond to $R^*=R_{300}$ and squares stand for $R^*=R_{max}$ 
giving better agreement with theory.  
Open symbols represent $T_{BKT}$ determined from the power-law behaviour of $I$-$V$ curves; 
closed symbols correspond to $T_{BKT}$ obtained from flux flow resistance data.  
The lines represent BMO functions \eqref{BMOfunc1} 
for different values of $\beta$ in \eqref{BMOfunc2}.  
}
\end{figure}

We now turn to the determination of $T_{BKT}$ in our samples.  
There exists two distinct methods for finding $T_{BKT}$ from transport measurements.  
The first utilizes power-law fits to the $I$-$V$ characteristics
of the film for which the switch from $V\propto I$ to $V\propto I^3$ 
occurs at the transition~\cite{HalperinNelson1979}.
Figures\,\ref{fig:VBKTexp}a,b show the typical sets of the 
$I$-$V$ curves for our samples at different temperatures 
near and below $T_{BKT}$ as log-log plots which indeed represent 
$V\propto I^{\alpha}$ behaviour, with $\alpha$
rapidly growing in a narrow temperature window, characteristic 
to the BKT transition (Fig.\,\ref{fig:VBKTexp}c).
The values of $T_{BKT}^{(I)}$ determined as temperature where $\alpha=3$ 
are presented in the Table 1.
Note that even at $T>T_{BKT}$ while $V\propto I$ at low currents, 
the $I$-$V$s become strongly nonlinear at 
elevated currents showing the characteristic rounding.  
In contrast to that at $T\leq T_{BKT}$ there is
an abrupt voltage jump terminating the power-law behaviour at a certain well defined current.
We attribute this jump to a heating instability in the low-temperature BKT phase, 
the nature of which along with the 
formation of the critical current in the BKT state will be discussed 
in a forthcoming publication~\cite{GurVinBat}.

The second technique for determining $T_{BKT}$ involves the use of flux flow resistance 
data~\cite{Minnhagen1981,MinnhagenRevModPhys,Fiory1983,HebardPaalanen1985,Simon1987}.  
Figure~\ref{fig:VBKTexp}d
shows a family of magnetoresistance isotherms typical for all our samples.  
With increase of temperature the $R(B)$ dependences progress 
from positive curvature through linear dependence to negative curvature.  
Below $T_{BKT}$ field-induced free vortices not only contribute to the resistance 
due to their own motion, but screen antivortices helping to dissociate vortex-antivortex pairs.  
This results in a superlinear 
response to the applied field.  
Above $T_{BKT}$ the vortex-antivortex pairs are unbound and all thermally 
induced vortices contribute to the resistance.
Field-induced vortices annihilate with some fraction of antivortices, 
effectively reducing the number of vortices participating in the flux-flow,
giving rise to a sublinear response to the applied field.  
The temperature of the linear response indicates thus $T_{BKT}$.  
The corresponding temperatures denoted as $T_{BKT}^{(B)}$ are listed in Table I.
Note that while $T_{BKT}^{(B)}$ and $T_{BKT}^{(I)}$ are very close to each other,
the former appears to be slightly lower than the corresponding values of $T_{BKT}^{(I)}$.

The summary of our results is presented in Fig.~\ref{fig:TBKTtoTcexp}, showing 
the ratio of $T_{BKT}/T_c$ vs. $R_{300}$ and $R_{max}$.  Irrespectively to the choice
of the resistance, the more disordered the film (i.e. the closer the film is to the SIT),
the farther apart $T_{BKT}$ from $T_c$ is. 
Using the dirty-limit formula which relates the two-dimensional
magnetic screening length to the normal-state resistance $R_N$,
Beasley, Mooij, and Orlando (BMO)~\cite{Beasley1979} (see also~\cite{HebardKotliar1989})
proposed the universal expression for ratio $T_{BKT}/T_c$
\begin{equation}
\frac{T_{BKT}}{T_c}f^{-1}\left(\frac{T_{BKT}}{T_c}\right)
=0.561\frac{\pi^3}{8}\left(\frac{\hbar}{e^2}\right)
\frac{1}{R_N}\,,
\label{BMOfunc1}
\end{equation}
\begin{equation}
f\left(\frac{T}{T_c}\right)
=\frac{\Delta(T)}{\Delta(0)}\tanh\left[\frac{\beta \Delta(T)T_c}{2\Delta(0)T}\right]\,,
\label{BMOfunc2}
\end{equation}
where $\Delta(T)$ is the temperature dependence of the superconducting gap and
parameter $\beta = \Delta(0)/(k_B T_c)$. The BCS theory predicts value $\beta=1.76$.
The BMO function for this value of $\beta$ is shown by the solid line 
in Fig.~\ref{fig:TBKTtoTcexp} and correctly describes the qualitative tendency of decreasing the 
$T_{BKT}/T_c$ ratio with increasing disorder. 
The quantitative comparison however encounters some problems.  The source 
of one of them is that since disordered films exhibit strong temperature dependence of the resistance
the choice of what value is to be taken as the normal-state resistance, 
$R_N$ in Eq.\,\eqref{BMOfunc1}, is not \textit{a priori} obvious, 
see the extensive discussion of the ambiguity of the experimental 
definition of $R_N$ in~\cite{HebardKotliar1989}.  
Indeed, we see that choosing $R_{300}$ as $R_N$, we obtain $T_{BKT}/T_c$ dropping much faster
that what is predicted by the BMO formula.  
At the same time taking $R_{max}$ as normal 
state resistance yields a result more close to theory, see Fig.~\ref{fig:TBKTtoTcexp}.
Another source of discrepancy is that the BMO function contains 
the ratio $\Delta(0)/T_c$.
As it is shown in ref.~\cite{STM_TiN}, where the TiN films close by the parameters to those
investigated in the present work were studied, this ratio is unusually large as compared to its BCS
value and grows on the approach to the SIT.  
The dashed lines in Fig.~\ref{fig:TBKTtoTcexp} show the BMO function
for elevated values of $\beta$ and demonstrate that choosing values of 
$\beta$ corresponding to increasing disorder
can improve the agreement between theory and experiment. 

In conclusion, we demonstrated that the temperature dependence of the resistance 
of quasi-two-dimensional superconducting films,
including its non-monotonic behaviour and the significant broadening of the transition 
is perfectly described by the theory of quantum contributions to conductivity.  
The analysis based on careful account of all contributions enabled a
precise determination of the superconducting transition temperature.  
We found that the transition to the global phase-coherent superconducting state occurs 
via the Berezinskii-Kosterlitz-Thouless transition, and that
the ratio $T_{BKT}/T_c$ follows the universal Beasley-Mooij-Orlando 
relation upon an appropriate choice of the normal 
state resistance and taking into account the non-BCS $\Delta(0)/T_c$ ratio in disordered films.

\begin{acknowledgments}
This research is supported by the Program ``Quantum Physics of Condensed Matter'' 
of the Russian Academy of Sciences, by the Russian Foundation for Basic Research 
(Grant No. 09-02-01205), and by the U.S. Department of Energy Office of Science 
under the Contract No. DE-AC02-06CH11357. 
\end{acknowledgments}

\end{document}